**Characterizing Large Strain Elasticity of Brittle Elastomeric Networks by Embedding them in a Soft Extensible Matrix**

*Etienne Ducrot[‡], Costantino Creton\**

Dr. E. Ducrot, Prof. C. Creton

Laboratoire Sciences et Ingénierie de la Matière Molle, ESPCI ParisTech, 10 rue Vauquelin, 75231 Paris Cedex 05, FRANCE

E-mail: costantino.creton@espci.fr

[‡]current address CSMR, NYU Physics, 4 Washington pl. 10003 New York, NY

**Abstract**

*We recently reported the general design and properties of new multiple network elastomers with an exceptional combination of stiffness, toughness and elasticity[1]. In the present paper, we report in more detail how the increase in strain at break resulting from the toughening can be used to provide great insight in the large strain properties of otherwise brittle acrylic well crosslinked networks. The networks have been prepared by sequences of polymerization and swelling with monomers. The parameters that have been varied are the nature of the base monomers and the degree of crosslinking of the first network. We characterize the small strain properties, equilibrium swelling and large strain properties in uniaxial tension. We show that the large strain properties of the multiple networks are quantitatively controlled by the large strain properties of the stretched first network which acts as a percolating filler, while the small and intermediate properties are controlled by the entanglement density which can be largely superior to that of homogeneous networks.*

**Introduction**

Elastomers are widely used because of their large reversible deformability up to strains of several hundred percent. Their elastic behavior stems from entropic elasticity of crosslinked flexible chains with a low glass transition temperature.[2] In the linear regime, the elastic modulus is typically given by the number of elastic chains per unit volume, which in the absence of a swelling solvent and for a homogeneously crosslinked system, is inversely proportional to the average molecular weight between crosslinks $M_x$.[3] Yet in uniaxial extension the majority of unfilled elastomers soften slightly relative to the linear behavior and then break in a brittle fashion well before the chains reach their maximum extensibility.[4] As a result, and with the notable exception of natural rubber, it is quite difficult to prepare an unfilled, well crosslinked elastomer, displaying a sufficiently high strain at break to characterize the distinctive strain stiffening in uniaxial elongation due to the finite extensibility of the chains.

This intrinsic limitation of conventionally crosslinked elastomers is due to the very localized nature of the breakage of bonds near the interfacial plane. In the absence of viscoelastic dissipation, the threshold fracture energy $\varGamma_0$ of the elastomers is quantitatively explained by the sequential fracture of highly stretched elastic strands in the plane of the interface as the crack propagates.[1, 5]

To avoid this damage localization process leading to early fracture, it is possible to modify the network architecture to create additional energy dissipation mechanisms. Many strategies have been tried to increase the toughness of unfilled elastomers by network design..[6] [7] [1, 8]

Our group has recently reported the successful design of multiple interpenetrating network elastomers through the sequential polymerization of acrylic monomers, in such a way that the network synthesized first becomes progressively isotropically stretched by the swelling while a second and third polymerization result in nearly unstretched and loosely crosslinked chains[1]]. When the elastomer is highly stretched or broken, the prestretched network chains break in the bulk of the elastomer well before the unstretched ones, significantly delaying the nucleation and propagation of a macroscopic crack. The best of the reported acrylic elastomers are very stiff (up to E = 4 MPa) and combine extensibilities of 2-300% with true stresses at break in excess of 30 MPa and values of fracture energies of 9000 J/m$^2$ while maintaining residual deformations immediately after unloading of only a few percent of strain. One of the key advantages of these interesting new materials is their very pronounced strain stiffening in uniaxial extension which now occurs before the material breaks. In some specific applications such as resistance to cavitation[9], or dielectric elastomers[10] one expects this feature to be very useful to fine tune the properties. Furthermore the characterization of the strain stiffening of the material may be directly related to the progressive stiffening of the prestretched polymer chains as they approach their limiting extensibility.

In the current paper we explore how, by changing the monomer composition and the degree of crosslinking used in the first network it is possible to control the mechanical properties of the final material and to characterize the strain stiffening of an otherwise brittle elastic polymer network.

## 2. Results
### 2.1. Composition, homogeneity, elasticity

Multiple networks elastomers have been prepared by sequential swelling and polymerization steps as described in the experimental section and the main synthesis parameters are given in **Tables 1** and **2**. The multiple network elastomers are composed of short and extended chains from the first network and long and nearly Gaussian chains from the second and third networks. All three networks are continuous throughout the sample. The measurement of the extractible shows that the proportion of free chains is negligible, (< 1 wt %) for every double network (DN) and triple network (TN). Different monomers have been used and result in different polymer chains connected together in a network: poly(methyl acrylate) (PMA), poly(ethyl acrylate) (PEA) and poly(butyl acrylate) (PBA). Samples are identified as $X_y[Z[Z]]$, with 'X' the first network monomer and 'y' its crosslinker concentration, 'Z' the monomer of the second/third network. The question of the miscibility, at the repeat unit level, between these different elastic chains needs to be addressed.

Dynamic Mechanical Analyzer measurements of G' and G" at $\dot{\varepsilon} = 1$ s$^{-1}$ as a function of temperature, show only one $T_g$ for DN and TN containing PMA and PEA (**Figure 1a**), in agreement with a non phase separated system. However, the DN composed of PBA and PMA clearly has two peaks in *G"* as a function of temperature suggesting phase separation. Those samples stay nevertheless perfectly transparent, indicating that the phase separation may occur at a molecular length scale as in microphase separated block copolymers.[11]

At T ~ 40 °C above the highest Tg, every multiple network behaves as a rubbery material in which the modulus increases linearly with the temperature and present a dissipative behavior characterized by a low value of tanδ as shown in **Figure 1b**. This remains the case for the microphase separated BA$_1$MA.

## 2.2. Equilibrium swelling of the first network

Immersed in a bath of solvent (the monomer methyl acrylate in our case), a polymer network swells to an equilibrium value where the loss in entropy due to chain stretching is compensated by the gain in entropy of mixing with the solvent molecules.[3] The swelling ratio is then governed by the crosslinker concentration of the network, its structure and the quality of the solvent. If the methyl acrylate monomers that swell the network are polymerized, the chain stretching of the first network occurring during the swelling is then frozen in place by the polymerization of the chains of the second network[12]. If such a DN is immersed again in a monomer bath, it will swell again until the loss in entropy due to the stretching of the 1$^{st}$ and 2$^{nd}$ network chains is compensated by the gain in entropy of mixing with the liquid monomer. Such a method gives a versatile way to control the volume fraction of the stiff network chains and has been used for hydrogels as well.[13]

However after the 3$^{rd}$ polymerization, there will be three chain populations, a highly stretched but very diluted population of 1$^{st}$ network chains, effectively constituting a stiff network, a proportion of lightly stretched 2$^{nd}$ network chains and a majority of unstretched 3$^{rd}$ network chains. **Figure 2** shows the equilibrium swelling of the PEA first network in the EA$_y$MA double and EA$_y$MAMA triple networks networks after polymerization as a function of the crosslinker concentration in the 1$^{st}$ network. Each swelling equilibrium defines the degree of isotropic prestretching $\lambda_0$ of the 1st network chains and those quantities for all samples tested in this study are summarized on **Table 3**.

## 2.3. Mechanical properties in uniaxial deformation

The elastic properties of simple crosslinked networks above their glass transition temperature can be typically tuned by changing the crosslink density. A higher concentration of crosslinker leads to a higher density of elastic chains and therefore to a higher elastic modulus. The effect of changing the crosslink density on the uniaxial stress-strain curves for three different SN of PEA is reported on **Figure 3a**. A side effect of increasing the modulus is typically however a strong decrease of the extensibility of the material which is consistent with the Lake-Thomas model[14].

Compared to these simple polyacrylate networks, our multiple networks present striking differences in mechanical behavior. As shown in **Figure 3b** for the case of the $EA_{0.5}$ family of samples, while the simple network of $EA_{05}$ and the second network of polymerized alone $MA_{2N}$ appear to follow Gaussian elasticity with no dramatic change in curvature, the sequentially polymerized DN $EA_{05}MA$ shows a pronounced stiffening at high strain and the TN $EA_{05}MAMA$ presents both a stiffening at intermediate strains and then a softening at high strain. Furthermore the initial modulus clearly increases after each polymerization.

For the example of **Figure 3b**, several mechanical properties are enhanced relative to the parent SN. The initial modulus after each polymerization step increases from E = 0.8 MPa for the SN to 1.5 and 2.4 MPa for the derived DN and TN respectively. Concurrently the elongation at break also increases from $\lambda_b$ = 1.5 to nearly $\lambda_b$ = 2.5 in DN and TN relative to the corresponding SN which is very brittle and poorly extensible (**Figure 3a**). The nominal stress at break is also highly enhanced by a factor of 8 and 28 relative to the SN for the DN and TN respectively. For the sake of comparison we also show a simple network ($MA_{bulk}$) of MA prepared to have a Young's modulus in the range of that of a TN. $MA_{bulk}$ is much more brittle than a TN with the same initial modulus, showing the effectiveness of incorporating prestretched chains in a matrix to combine stiffness and toughness. It should be noted that those values correspond to the highest values obtained from each type of sample.

**Figure 3a,c** and **d** shows the effect of the crosslinker concentration of the stiff network in the mechanical properties of single networks and derived multiple networks. As discussed above, while the initial modulus of the SN increases with the amount of crosslinker in the material (**Table 4**), the crosslinker concentration in the first network has also an important effect on the properties of the DN and TN made from the respective SN's. By adding more crosslinks in the first network, we increase the initial modulus of the DN and TN and more generally the stress level for all $\lambda$ is very significantly increased. Secondly, we observe clearly that the onset of strain stiffening occurs at a lower extension for the DN and TN made from the more densely crosslinked first networks.

A second parameter that can be changed is the monomer used for the synthesis of the 1$^{st}$ network. **Figure 4** shows the stress-strain curves of three different first networks that have been synthesized from MA, EA and BA with the same weight fraction of crosslinker, and their corresponding double networks, with MA as a second monomer. Although the targeted average number of monomers between crosslinks is different, the targeted average molecular weight between crosslinks, and hence the elastic modulus are the same. The three DN all contain 30 wt % of first network (**Table 3**). In uniaxial extension, the three materials have indeed a very similar initial modulus in small strain (2 MPa) but their stress-strain curves diverge for $\lambda > 1.4$ and interestingly, the weaker first network gives the stronger double network in term of stress at break. The stiffening is observed for lower $\lambda$ in the BA$_1$MA sample than in the MA$_1$MA, the stiffening of which is very limited. The nominal stress at break is multiplied by a factor of almost 5 between those two samples. $EA_1MA$ being between the two extreme.

Finally, multiple networks elastomers of pure PEA have also been prepared starting from a first network of $EA_{0.5}$. Mechanical tests were performed at 20 °C in order to be around 40 °C higher than the $T_g$ of the pure EA networks. The uniaxial stress-strain curves for this series are shown in **Figure 6** in comparison with the equivalent series but with MA as second and third monomer. The general aspect of the stress-strain curves between the two series is very similar except that the initial modulus and stress level are generally lower for fully EA based networks.

*3. Discussion*

While the increase in the mechanical properties of these interpenetrated and very elastic networks is remarkable, it is interesting to discuss in more detail how to control their properties by changing their composition (monomers, crosslinker)

**3.1. Influence of the choice of monomer**

In double network hydrogels, the nature of the two monomers has to be carefully chosen[13] to achieve a high mechanical reinforcement. Typically permanently charged polymers are used in the first network to reach equilibrium swelling ratios of the order of 10 and tough gels can be obtained in only two polymerization steps.

In the case of multiple network elastomers the need for a relatively low $T_g$ and an elastic behavior precludes the use of very polar monomers able to hydrogen bond strongly or to form dipole-dipole interactions. Given these restrictions, changing the monomer used to synthesize the stiff network appears to have a limited influence on the mechanical properties. The networks $MA_1$, $EA_1$ and $BA_1$ have the same average molecular weight between crosslinks $M_x$ of 7500 g.mol$^{-1}$ (from Table 4), but a number of repeat units $N_x$ between crosslinks which decreases with the molar mass of the monomer: $N_x$= 86 for $MA_1$, 76 for $EA_1$ and 59 for $BA_1$. Chains between crosslinks have therefore less backbone bonds in $BA_1$ than in $MA_1$ and since the maximum extensibility of a chain $\lambda_{max}$ scales with $N_x^{1/2}$ one expects $\lambda_{max}$ to scale with $M_0^{-1/2}$ at a fixed value of $M_x$. Because the prestretching of chains $\lambda_0$ is identical for the three DN samples of **Figure 4b**, the stress-strain curves of **Figure 4b** replotted as a function of $\lambda N_x^{-1/2}$ should show a strain hardening at the same relative extension. **Figure 5** shows that it is not the case for the $BA_1MA$ which can stretch more relative to its unstretched length than the $EA_1MA$ and $MA_1MA$ network. This may be due to the heterogeneous structure of the elastomer resulting from the microphase separation(**Table 3**). Note that this strain stiffening at higher values of l does appear to modify or reduce the reinforcement mechanism and the stress at break is higher. In elastomers, the multiple network effect is therefore still effective even if the two networks are not fully miscible. The main point is to introduce prestretched chains in the system and a large contrast in crosslinker concentration between the networks.

The comparison between the multiple networks made from an identical 1st network but where the unstretched networks are made from MA or EA is instructive. It shows first that a difference in monomer composition between the stiff and the soft networks is not needed for mechanical reinforcement in multiple network elastomers. However PMA and PEA melts do not have the same average molecular weight between entanglements. Although detailed measurements on PMA are difficult to find there is evidence that EA is less entangled than MA ($M_e$(EA) = 13 kg.mol$^{-1}$ [15] and $M_e$(MA) = 11 kg.mol$^{-1}$ [16]) and this may explain the differences in stress level observed. Those simple experiments demonstrate that in multiple network elastomers different monomers can be used to fine tune the properties as long as they can swell the first network sufficiently.

## 3.2. Stretching of chains in the stiff network due to swelling

In the swollen state or in the DN and TN, we assume that, even in the absence of any macroscopic deformation, the chains of the first networks are isotropically stretched by a factor $\lambda_0$ relative to the dry state as defined on **Equation (1)** by the ratio of thickness of the final network ($h$) and the thickness of the corresponding first network ($h_{SN}$). For DN, the reported values of $\lambda_0$ on **Figure 7** stay below 1.7. Given that the 1$^{st}$ network is synthesized in solvent condition (in the presence of 50% toluene) and then subsequently dried, chains are slightly supercoiled in the preparation condition they are then only slightly extended compared to their Gaussian state once swollen. Once the polymerization of the 2$^{nd}$ monomer has occurred inside the 1$^{st}$ network, the entropy of mixing between monomer and network disappears and these DN can again be swollen with monomer. This 2$^{nd}$ swelling step leads then to a higher stretching of the first network, up to a $\lambda_0$ of 2.5 and a volume fraction of 1$^{st}$ network well below 10%. Such a high level of prestretching is not achievable in one step with neutral polymers.

$$\lambda_0 = \frac{h}{h_{SN}} \tag{1}$$

The degree of prestretching (and the volume fraction) of 1$^{st}$ network chains can be controlled through the number of swelling/polymerization steps or by the crosslinker concentration in the first network. An increase in crosslinker concentration gives a network with shorter chains between crosslinks and the equilibrium swelling decreases.

It should be noted that Yoo et al.[17] reported no significant mechanical enhancement of the toughness of silicone double networks in a case where the loose network is first prepared and then swollen by short oligomers subsequently crosslinked. Following this path, they did not introduce short prestretched chains in their system and the volume fraction of unstretched short chains was larger than that of the long chains. The prestretching of a fraction of chains in the final multiple networks appears therefore to be important to enhance the mechanical properties.

### 3.3. Origin of the initial modulus

An interesting feature of both DN and TN is the significant increase in modulus which is observed. For every family of multiple networks (DN or TN), the relation between the modulus E and the concentration of chemical crosslinker in the first network seems to be linear (**Figure 8**). However the absolute values are much higher for the DN series and then the TN series than for the three single networks and the slope is steeper when increasing the number of polymerizations, meaning that the impact of the crosslinker concentration in the first network becomes dominant after three steps of polymerization.

Multiple networks are composed of prestretched chains of the first network at $\lambda_0$, and loosely crosslinked chains on second/third networks which are entangled together and loosely connected. In the TN there are more precisely three populations of chains: extended chains of the first network, lightly stretched entangled chains of the second network (although the very small amount of crosslinker introduced in the second network suggests that the $2^{nd}$ network chains can relax during swelling and are only lightly stretched. An estimate of the contribution to the modulus from the prestretched chains of the first network can be made using **Equation (2)**.[28]

$$E_{1^{st}} = 3nRT \frac{\overline{R}^2}{\overline{R_0}^2} = \phi_{1^{st}} \times E_{1^{st}} \times (\lambda_0)^2 \tag{2}$$

with $\phi_{1^{st}}$ the volume fraction of first network, $E_{1^{st}}$ the initial modulus of the first network alone and $\lambda_0$ the isotropic elongation of the first network in the final DN or TN due to swelling. We now propose two ways to take into account the contributions from additional chains of the second and third networks.

If the average molecular weight between entanglements of the second network is significantly lower than the average molecular weight between crosslinks of the first network, the second and third networks should simply contribute to the modulus in proportion to their weight fraction $\phi_{2^{nd}}$ and $\phi_{3^{rd}}$. In TN, the prestretching of chains of the second network has to be taken into account by introducing an elongation of the second network $\lambda_0^{2^{nd}}$. The final formula for the estimate of the modulus of DN and TN for this case is given in **Equation (3)**. For every estimate the modulus of the second network alone $E_{2^{nd}}$, mainly due to the entanglement structure, has been taken at 0.9 MPa.

$$E_{calc} = \phi_{1^{st}} \times E_{1^{st}} \times (\lambda_0)^2 + \phi_{2^{nd}} \times E_{2^{nd}} \times \left(\lambda_0^{2^{nd}}\right)^2 + \phi_{3^{rd}} \times E_{2^{nd}} \tag{3}$$

If on the other hand the crosslinks of the first network act as topological constraints and delimit a confining distance between additional entanglements for the second and third network, the expression for the modulus must be modified as presented in **Equation (4)** for DN. The expression for TN is more delicate to establish and is not treated here, it should include the topological constraints from chains and crosslinks of the second network, as well as its stretching due to swelling.

$$E_{calc} = \phi_{1^{st}} \times E_{1^{st}} \times (\lambda_0)^2 + \phi_{2^{nd}} \times \left( E_{2^{nd}} + \frac{3\rho RT}{M_{mesh}} \right) \quad (4)$$

where $M_{mesh}$ is the average molar mass of the gaussian chains of the second network between two stretched strands of 1$^{st}$ network and can be expressed by **Equation (5)** in the case of a DN, taking into account the equilibrium degree of swelling and the difference in monomer type. For that we consider $M_x^{1^{st}}$ the average weight between crosslinks in the unswollen first network of PEA. In the DN, due to swelling, the mesh is isotropically stretched by a factor $\lambda_0$. The distance $\mathcal{R}$ between crosslinks of the first network scale as $\mathcal{R} \sim \left(M_x^{1^{st}}\right)^{1/2} \lambda_0$, leading to the expression of $M_{mesh}$ given on **Equation (5)**.

$$M_{mesh} = \left( \frac{M_{EA}}{M_{MA}} \right) M_x^{1^{st}} \lambda_0^2 \quad (5)$$

Estimates of the initial modulus obtained from **Equations (3)** and **(4)** are reported on **Figure 8** with the experimental modulus. The estimate of the modulus from **Equation (3)** predicts well EA$_{0.5}$MA, but fails for all other samples, showing that the assumption of simple additivity of entanglements from the two networks is incorrect for highly crosslinked first networks. **Equation (4)**, assumes effectively that the crosslinks of the first network control the entanglement density of the second network in an analogous way to a fixed network of topological obstacles controlling the diffusion of long chains through.[32] The predicted values of E from **Equation (4)** are in much better agreement with the experimental data than the previous model but still underestimate the modulus for EA$_2$MA. In this last sample the mesh size of the swollen first network corresponds to an equivalent PMA chain of 5500 g.mol$^{-1}$, significantly smaller than the average weight between entanglements for pure PMA. Some additional entanglements between polymer chains of the two networks themselves have to be added for a better description, and not only a contribution from crosslinks of the first network.

From those simple models in DN we can estimate that for loosely crosslinked first networks (i.e. $M_x^{1st} \gg M_e^{2nd}$), the key parameter to explain the modulus is the level of prestretching of the chains of the first network. Then for more crosslinked first networks (i.e. $M_x^{1st} \sim or < M_e^{2nd}$) the crosslinks and chains of the first network form topological constraints for second network chains and are responsible for the increase in modulus by increasing the density of entangled chains. For TN, the high dilution of stretched chains also dilutes their contribution to the modulus (1$^{st}$ model ineffective) and the presence of three entangled networks with different degrees of prestretching and some connections between networks, make the second model more difficult to use.

In the supplementary information we discuss the impact of transfer reactions during the polymerization of acrylates in the bulk. As discussed for hydrogels[18], the remaining unreacted vinyl groups after the polymerization of the 1$^{st}$ network and the normally occurring transfer reactions to the polymer will create some connecting points between the networks and so participate to the increase in modulus. A direct estimate of the connections between the two networks is difficult after the polymerization of the second and third networks. We estimated this contribution to be around 0.25 MPa, it is nevertheless too low to account for the large observed increase in modulus for the DN made from highly crosslinked first networks.

## 3.4. Interpretation of the stiffening at high strain

Double and triple networks show a marked stiffening in large strain in uniaxial extension and a high stress at break. For simple networks, samples break at too low an extension ratio to show stiffening and the second network of PMA alone (MA$_{2N}$) may present stiffening but at a very high elongation which is not accessible with our experimental setup (i.e. $\lambda > 5.5$). For the DN, the onset of stiffening, which corresponds to the abrupt change in slope of the stress/strain curve, occurs before an elongation $\lambda$ of 2, and even at lower values of $\lambda$ for the TN.

In unfilled rubber-like materials, the phenomenon of strain stiffening is due either to the fact that the chains forming the network are stretched close to their finite extensibility[19] or to strain induced crystallization[20] or strain induced clustering[21]. This finite extensibility has been discussed theoretically at the chain level[19] but although several empirical models exist[22, 23] there is currently no molecular model that captures quantitatively the effect of non-gaussian elasticity in a network[24, 25]. One of the reasons is that this regime is often inaccessible experimentally for a macroscopic material because of premature fracture and models can only progress if more systematic data is available.

Acrylates do not self-associate by hydrogen bonding, and cannot crystallize so the observed strain stiffening in multiple networks has to be due finite extensibility of the shorter and already prestretched 1$^{st}$ network chains in the system. To confirm this hypothesis we will analyze the data of a family of stress/strain curves of EA$_{0.5}$ based multiple networks reported on **Figure 9**. If we assume that chains of the 1$^{st}$ network are stretched affinely with the degree of swelling in the final system (**Table 5**) we can define a degree of stretching at preparation $\lambda_0$ and a corrected elongation $\lambda_{cor}$ defined by **Equation (6)**. In the multiple networks being stretched uniaxially, it corresponds to the real elongation of the chains of the first network in the direction of stretching relative to its end to end distance in the dry first network.

$$\lambda_{cor} = \lambda \cdot \lambda_0 \qquad (6)$$

Now that the elongation has been rescaled by the level of prestretching $\lambda_0$ (**Figure 9**), the stress $\sigma(\lambda_{cor})$ of the DN and TN needs also to be corrected to correspond to the stress of the SN alone at that value of $\lambda$. To do so, we rescaled the reduced stress (defined as $\sigma_R = (\sigma_N + \sigma_C)/(\lambda - 1/\lambda^2)$) before the strain stiffening to that of the first network, with $\sigma_C$ a translation constant. We can then fit the master curve obtained by rescaling the stress strain data. with an empirical strain stiffening model inspired by the Arruda and Boyce equation and modified with Mooney parameters[23, 26] to incorporate the softening at intermediate strain (**Equation (7)**).

$$\sigma_{Mooney} = \frac{2. C_1 \cdot \left(1 + \frac{C_2}{C_1 \cdot \lambda_{cor}}\right)}{1 - \frac{\lambda_{cor}^2}{\lambda_{limit}^2}} \qquad (7)$$

with $C_1$, $C_2$ the Mooney-Rivlin parameters and $\lambda_{limit}$ the limiting extensibility of the chains obtained from the fit. Once the master curve is constructed, $C_1$ and $C_2$ are fitted freely on the SN curve, $\lambda_{limit}$ is adjusted manually to set the stiffening at the right elongation defined by the rescaled DN and TN curves and an example for the $EA_{0.5}$ multiple networks family is shown in **Figure 9**. The functional form of the stiffening of the double network as a function of the corrected stretch is well described without adjustable parameters by the fit of the corresponding first network to **Equation (7)**. This experimental evidence shows that the stiffening is indeed controlled by the final extensibility of the minority chains of the first network. The stiffening of the triple network is also well described for $\lambda_{cor} < 4$. At higher elongation, the stress of the TN diverges however from the master curve suggesting a damage mechanism due to internal bond breakage[1]. which is not the focus of the present paper. The same master curve construction was also applied to $EA_1$ and $EA_2$ based multiple networks with the same success,. Some representative master curves and best fits of **Equation (7)** are shown in **Figure 10**, and presented in **Table 5**.

As observed in the $EA_{0.5}$ family of samples, the initial modulus of the DN and TN is not well described by the model (**Figure 10**) confirming that the initial modulus is not purely the result of the prestretching of the chains of the first network. The stiffening however is well described by the model and directly relates to the finite extensibility of the minority chains of the first network. As can be seen in **Table 4**, the maximum extensibility of the chains of the first network as fitted by **Equation (7)**, decreases with increasing crosslinker level.

### 3.5. Maximum extensibility of chains of the first network

It is also in principle possible to estimate the maximum average extensibility of the first network chains from their elastic modulus $E$. From the number density of elastic chains obtained from the modulus, it is possible to extract the average number of C-C bonds per network chain. Then using **Equation (8)** derived from the Kuhn theory[3], and knowing $C_\infty$ for PEA, PMA or PBA, the theoretical maximum of extensibility of the chains $\lambda_{max}$ can be evaluated.

$$\lambda_{max}^2 = \frac{\mathcal{N}_{C-C} \cos\left(\frac{\theta}{2}\right)^2}{C_\infty} \tag{8}$$

Surprisingly for every family of samples, **Figure 11** shows a very good correlation between $\lambda_{max}$, estimated from the Young's modulus of the first network, and $\lambda_{limit}$, estimated from the stiffening parameter of **Equation (7)**. To the best of our knowledge this quantitative agreement is remarkable. This result has several interesting implications. First of all this correspondence suggests that the slightly supercoiled chains formed during the first polymerization in the presence of toluene, are nearly unentangled and that the Gaussian chains of the other two networks do not prevent full extension of the chain. In those conditions one can argue that the prestreched chains carry most of the load and that the stiffening measured on the macroscopic sample directly reflects the stiffening of the chains themselves.

More generally, these results suggest that swelling a network in a polymerizable monomer is a good method to characterize the stress-strain curve of the chains composing this network all the way to their fracture. This information can then be used to understand better fracture processes in polymers where stretches at the crack tip greatly exceed the values that can be measured in uniaxial extension. Furthermore the synthesis method presented here, in addition of producing tough materials, is a very good way to study the non-Gaussian elasticity of the polymer chains and could be used for much more model first networks as was recently done for hydrogels[27].

## 4. Conclusion

A series of multiple interpenetrating network elastomers have been successfully synthesized with sequential swelling/polymerization steps leading to interpenetrated multiple networks. This synthesis method leading to elastic tough elastomers can also be used to study the large strain behavior of simple networks that would otherwise break at low strain. We show that the stiffening at high strain in multiple networks can not only be tuned but also be predicted from the crosslinker concentration of the first network and its degree of swelling in the final sample. One can now furthermore use this simple and robust strategy to design and prepare synthetic rubbers with fine tuned nonlinear elastic properties involving significant strain stiffening. We believe that this strategy is universal and does not depend on polymer type as long as the material is in the rubbery state.

## 5. Experimental Section

*Materials:* The monomers, methyl acrylate (MA), ethyl acrylate (EA) and butyl acrylate (BA) and the crosslinker 1,4-butanediol diacrylate (BDA) were purified over a column of activated alumina to remove the inhibitor. The UV initiator, 2-hydroxy-2-methylpropiophenone (HMP) was used as received. Anhydrous toluene and cyclohexane were used as received as solvents. All reagents were purchased from Aldrich.

*Synthesis:* Both double and triple network elastomers were prepared by a multiple step free radical polymerization procedure starting from identical first networks.

*Preparation of the single network:* The stiff networks were prepared first by UV free radical polymerization of a solution of monomer (MA, EA or BA), BDA as crosslinker (1.45

– 5.81 mol % relative to monomer) and HMP as a UV initiator (1 – 1.49 mol % relative to monomer). The reactants were dissolved in anhydrous toluene as solvent at 50 wt %. All the reactants initially deoxygenated and stored in the glove box were mixed for a short time and poured in a 1 mm thick glass mold. The reaction was initiated by UV light (Vilbert Lourmat VL-215.L lamp, 365 nm, 10 µW/cm²) and left to proceed for 90 minutes. The sample was then taken out of the glove box and out of the mold, weighed and immersed in solvent baths for a week to remove every soluble components, unreacted species or free chains. The solvent mixture was changed every day with a gradual degradation of its quality to induce a slow deswelling of the sample. Samples were finally dried under vacuum at 80 °C overnight and then weighed and stored at room temperature for later use. Those samples will be referred as single networks (SN) and noted $X_y$ with 'X' referring to the monomer type (EA, MA, BA) and 'y' to the concentration of crosslinker.

*Preparation of DN and TN:* Starting from a first network (SN), double networks (containing a stiff network embedded in a loose one) were prepared following a swelling and polymerization sequence. A piece of first network was swollen in a bath composed of the second monomer (EA or MA), BDA as crosslinker (0.01 mol %) and HMP as UV initiator (0.01 mol %) as reported on **Table 2**. The swelling step was performed in a sealed plastic box in the glove box with oxygen-free reactants. Once swollen at equilibrium, after two hours, the sample was carefully extracted from the monomer bath, gently wiped to eliminate the excess of monomer, and placed between siliconized PET sheets and glass plates. The polymerization was initiated and completed via a two hours UV irradiation. The mold was then taken out of the glove box, and the sample was removed, weighed and dried under vacuum at 80 °C over night. It was finally stored at room temperature until later use. Those samples will be referred to as double networks (DN) and noted $X_yZ$ with '$X_y$' referring to the first network and 'Z' to the second monomer (EA or MA).

To reduce further the volume fraction of the 1st network, the same sequence of swelling/polymerization was repeated starting from double networks, the composition of the swelling bath being kept identical. Once again, after two hours of swelling, the sample was at equilibrium. The swollen double network was sealed between PET sheets and tightened between glass plates. The third network was polymerized inside the first two networks via a two hours exposure to the UV. The sample was then dried under vacuum at 80 °C overnight and stored at room temperature until later use. Those samples will be referred to as triple networks (TN) and noted $X_yZZ$ with '$X_y$' referring to the first network and 'Z' to the second and third monomer (EA or MA). It should be note that those network contain a stiff network (crosslinked and swollen), and two quite loose networks (very loosely crosslinked and almost unstretched). Transfer reactions between networks certainly occur during the second and third polymerizations so that these networks are loosely connected by covalent bonds.

*Preparation of the networks without solvent:* As references, loosely crosslinked networks of EA and MA were prepared alone in solvent free conditions by UV-initiated polymerization (HMP as UV initiator) between glass plates and a subsequent overnight drying step under vacuum at 80 °C to remove unreacted monomers. $EA_{2N}$ and $MA_{2N}$ prepared with 0.01 mol % of BDA are exact replicas of the second/third network but prepared alone (compositions in **Table 2**). $MA_{bulk}$ prepared with 2.5 mol % of BDA has been designed to present an initial modulus comparable to a TN.

*Characterizations:*

*Weight fraction of first network:* To determine the composition of the final DN, we used the weight and the thickness of the samples. Before swelling, the thickness ($h_{SN}$) and the weight ($m_{SN}$) of the first network were measured. The same characteristics were collected from the final DN and TN (h and m). The weight fraction ($\phi_{wt}$) of first network in the final multiple network was determined with **Equation (9).** The swelling ratio of the first network ($Q_{vol}$) was estimated from thicknesses (**Equation (10)**).

$$\phi_{wt} = \frac{m_{SN}}{m} \tag{9}$$

$$Q_{vol} = (\frac{h}{h_{SN}})^3 \tag{10}$$

*Extractible part:* A solvent extraction of free chains in DN and TN was performed to quantify the extractible part. A piece of material was weighed and then immersed in a large volume of toluene for two weeks. After drying under vacuum, the sample was weighed again. The weight loss corresponds to the amount of extracted free chains.

*Dynamic mechanical analysis (DMA):* DMA experiments were performed in tension mode on a Q800 DMA testing machine (TA instruments) on strips of material with a length close to 15 mm, a width of 4.9 mm and a thickness between 0.6 and 2.5 mm depending on the sample. A tensile preload force of 0.1 N was applied to the sample in order to avoid any buckling during the experiment. From this steady state, small oscillations were applied at 0.1 % of strain. Every test was performed at 1 Hz and 1 °C/min, from -90 °C or -50°C to 80 °C. The Tg of the material is taken at the maximum of the loss modulus.

*Mechanical tests:* Mechanical tests were performed on a standard tensile Instron machine, model 5565, using a 100 N load cell and an environmental chamber to precisely control the temperature (25 °C or 60 °C). A video extensometer gave a local measurement of the stretch $\lambda = l/l_0$ where $l_0$ is the initial gauge length. The relative uncertainty of the measurements given by the load cell and the video extensometer are respectively 0.1 % in the range of 0 to 100 N and 0.11 % at the full scale of 120 mm. Specimens were cut into a dumbbell shape using a normalized cutter (central part: length 20 mm, cross-section 4 mm and thickness 0.6 – 2.5 mm depending on the sample). Uniaxial tensile tests from small to large strain were performed at a constant velocity of the crosshead of 500 µm.s$^{-1}$ and the typical initial strain rate on the central part of the sample was around 0.04 s$^{-1}$. Using the nominal stress and strain, the Young's modulus was calculated at a strain of less than 10 %. For single networks, the classical theory of rubber elasticity was applied to determine the average molecular weight between crosslinks $M_x$.[3]

**Acknowledgements**
We gratefully acknowledge the financial support of DSM Royal B.V. and numerous fruitful discussions with Markus Bulters, Paul Steeman and Meredith Wiseman of DSM Ahead in Geleen. We also thank Hugh Brown for his insightful comments and encouragements on this project and Jian Ping Gong for her inspiration.

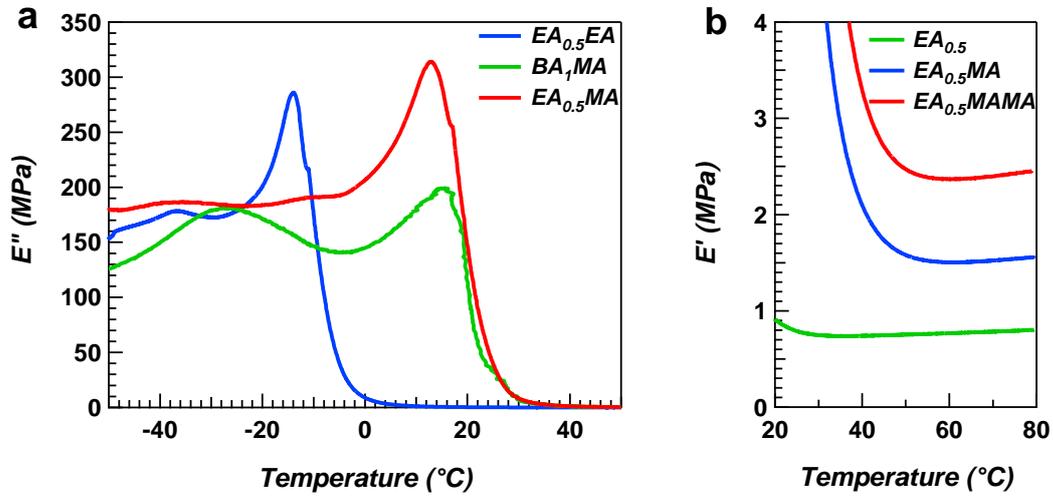

**Figure 1**: a) Loss modulus E'' as a function of temperature measured at $\dot{\varepsilon} = 1$ s$^{-1}$. Storage modulus measured by DMA at high temperature for the EA$_{0.5}$[MA[MA]] family of samples showing the elastomeric behavior for T > T$_g$

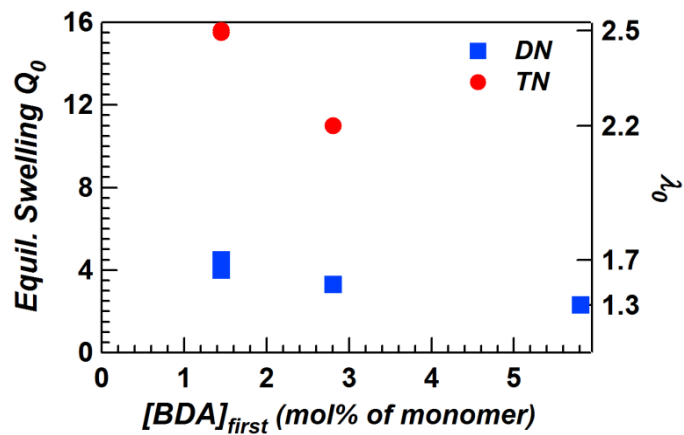

**Figure 2**: Equilibrium swelling ratio of the first network Q$_0$ and level of prestretching $\lambda_0$ as a function of the crosslinker concentration in DN and TN; [BDA]$_{first}$ being the concentration of crosslinker - 1,4-butanediol diacrylate - in the first network.

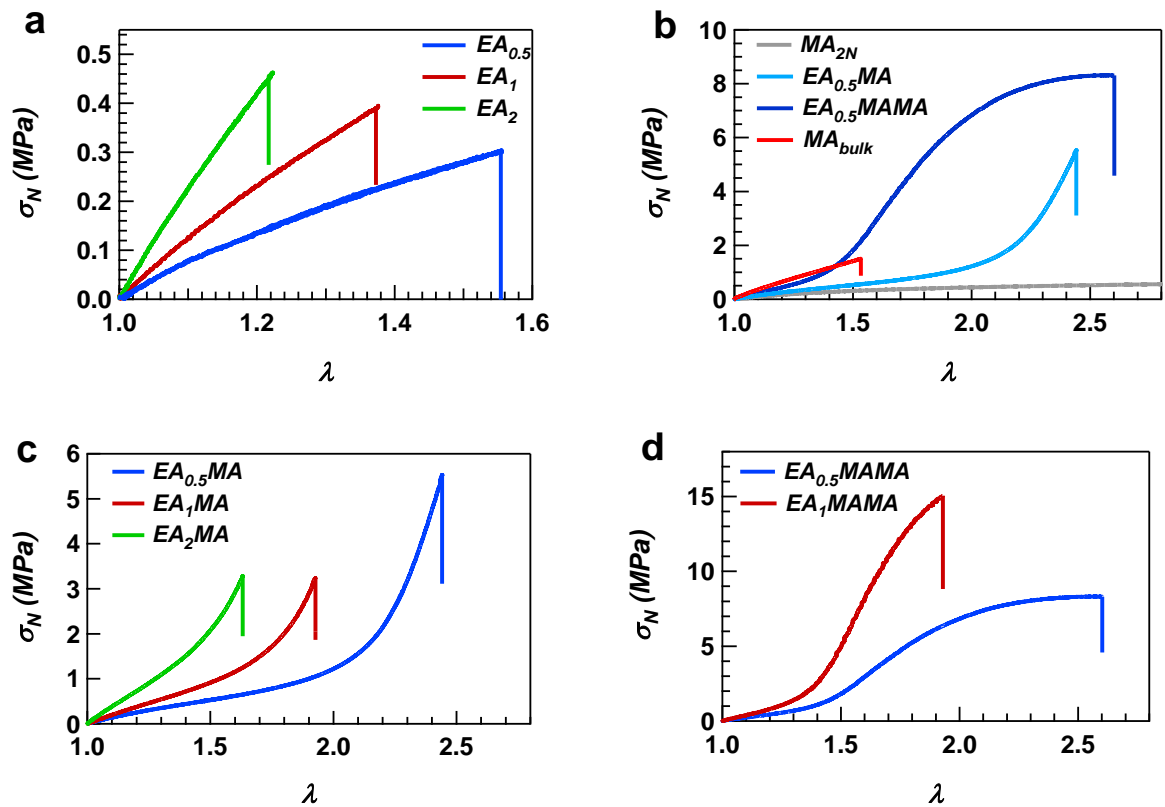

**Figure 3**: Stress/Strain curves at 60°C a) Simple networks of PEA with various crosslinker concentration; b) Comparison between $EA_x[MA[MA]]$ multiple networks (in blue), MA second network alone $MA_{2N}$ (in grey) and simple network of methyl acrylate prepared in bulk with 2.5 mol % of crosslinker $MA_{bulk}$ (in red). c) DN and d) TN with MA as second monomer prepared from SN corresponding to a);

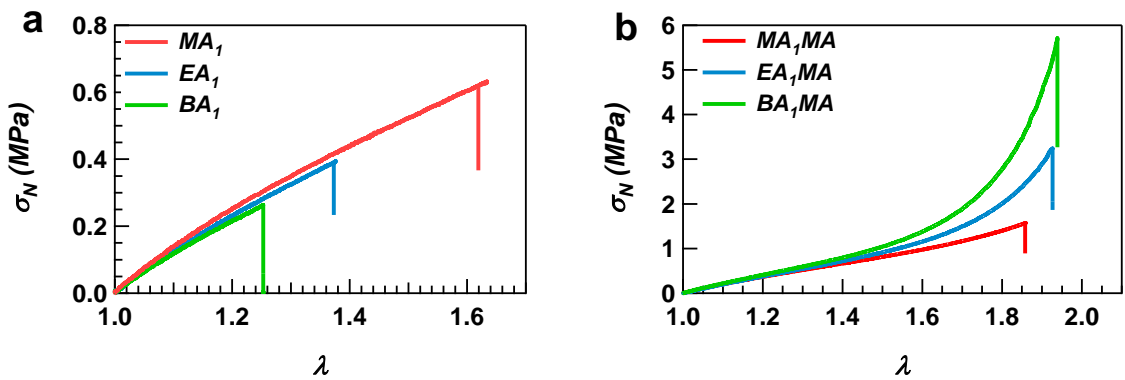

**Figure 4**: Monomer variation for the first network; corresponding Strain/Stress curves at 60°C for a) first networks and b) double networks with MA as second monomer

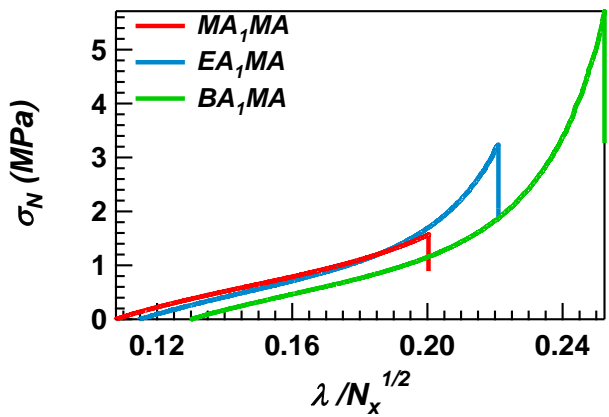

**Figure 5:** Data of **Figure 4** replotted as a function of $\lambda N_x^{-1/2}$.

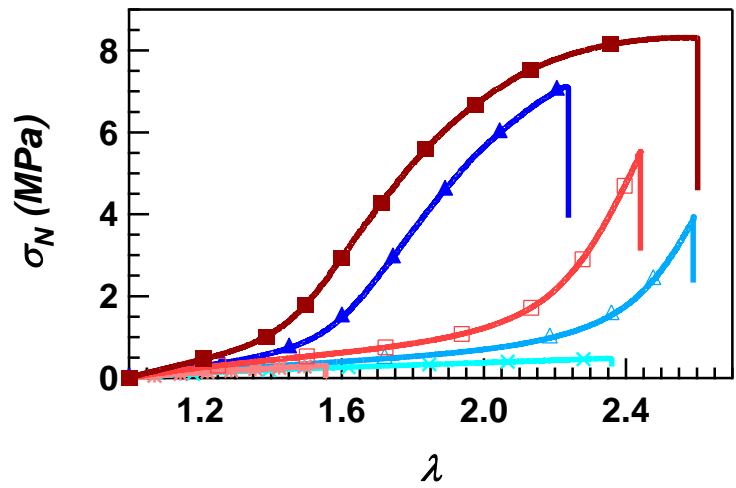

**Figure 6**: Effect on the second\third monomer on multiple networks 40 °C higher than the $T_g$; $EA_{0.5}$ as first network; (Red) MA as second/third monomer at 60 °C; (Blue) EA as second/third monomer at 20 °C; Cross for single networks, open symbols for double networks and full symbols for triple networks.

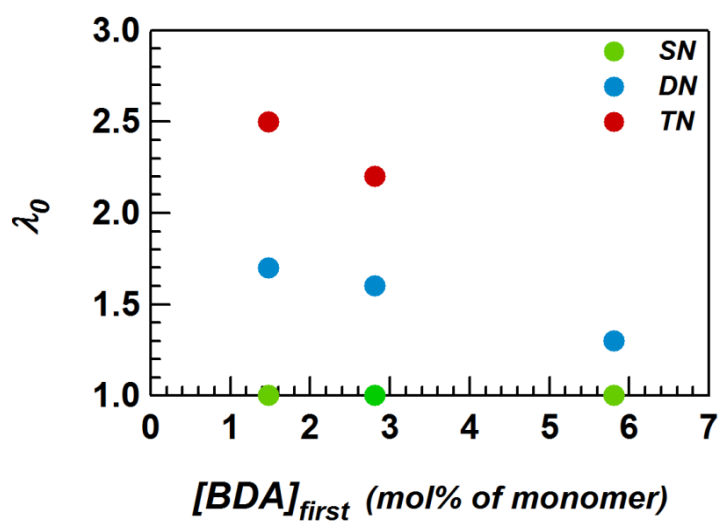

**Figure 7**: Swelling elongations or prestretch of chains of the first network ($\lambda_0$) versus the concentration of crosslinker in the first network for multiple networks prepared from $EA_y$ first networks and MA as second/third monomer.

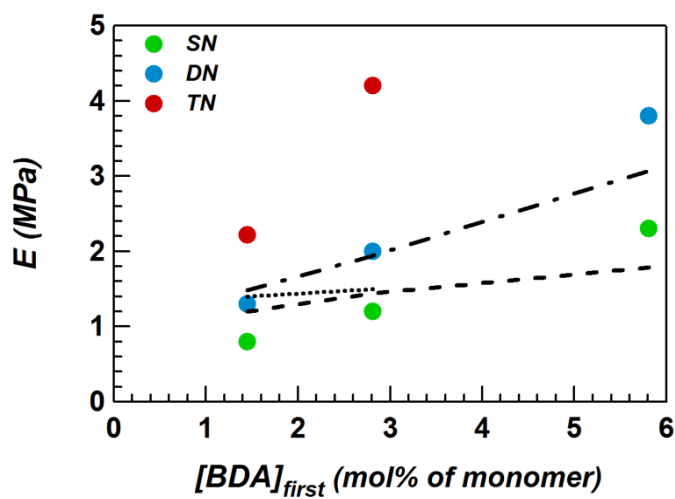

**Figure 8**: Initial modulus at 60°C versus concentration of chemical crosslinker in the first network for single and multiple networks $EA_y[MA[MA]]$; --- and … estimation respectively of DN and TN modulus from Equation 3; -.-.-.- estimation of DN modulus from Equation 4

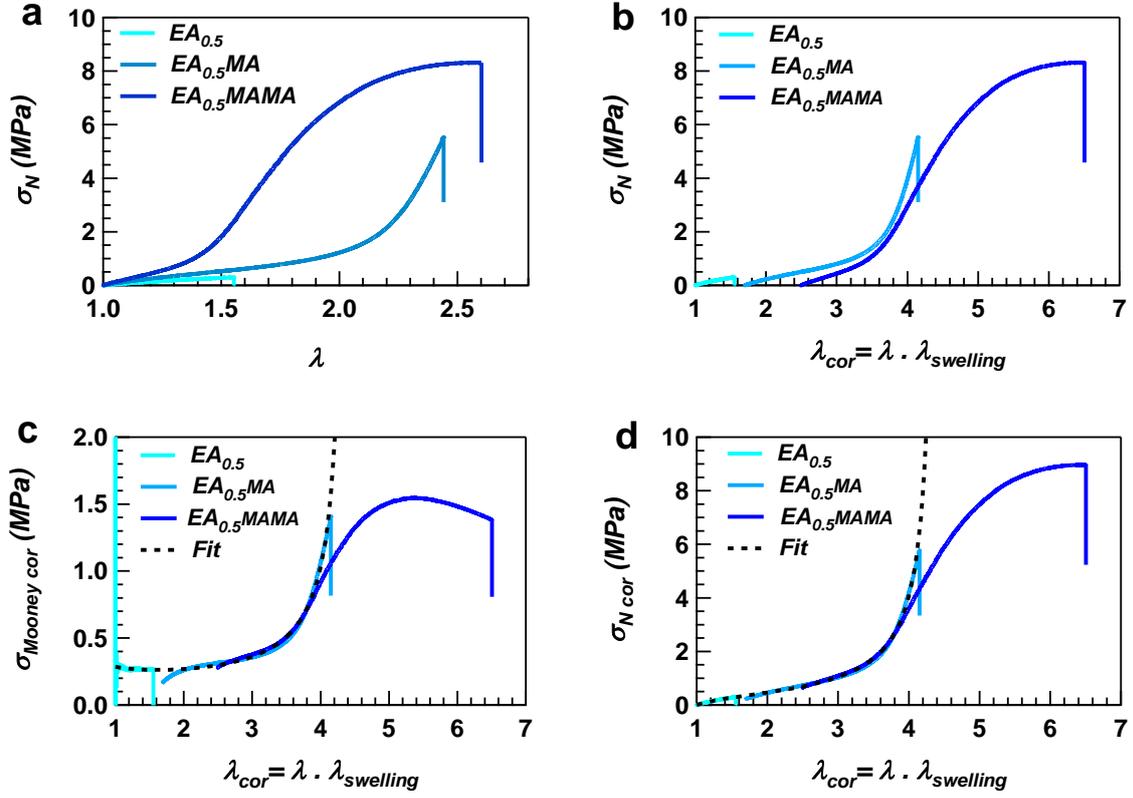

**Figure 9**: Construction of the master curve and modified Arruda-Boyce fits for the EA$_{0.5}$[MA[MA]] family of multiple network at 60°C; a) Initial Stress/Strain curves; b) After rescaling of the strain with the swelling of the first network; c) Corrected Mooney Stress ($\sigma_{Mooney\ cor}=(\sigma_N+C^{te})/(\lambda_{cor} - \lambda_{cor}^{-2})$) versus corrected Strain; d) Corrected Stress ($\sigma_{N\ Cor} = \sigma_N+C^{te}$) versus corrected Strain; dashed lines modified Arruda-Boyce best fit

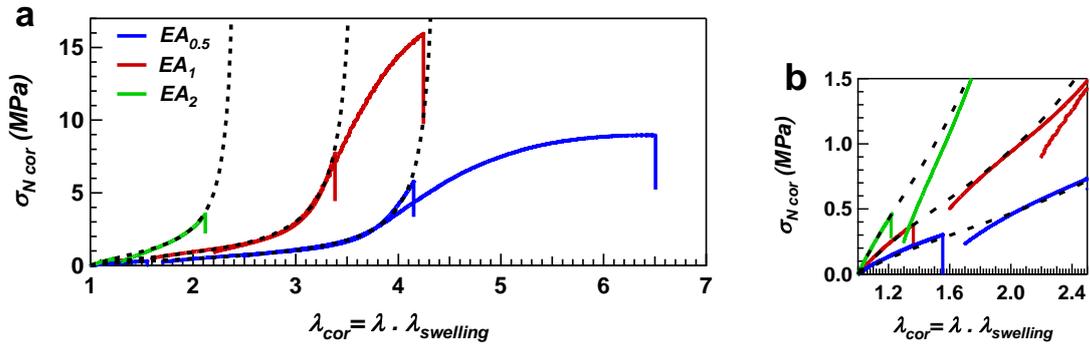

**Figure 10**: a) Master curves for $EA_{0.5}$, $EA_1$ and $EA_2$ based multiple networks; in dashed lines best fits of the modified Arruda-Boyce equation; b) Zoom at small strain

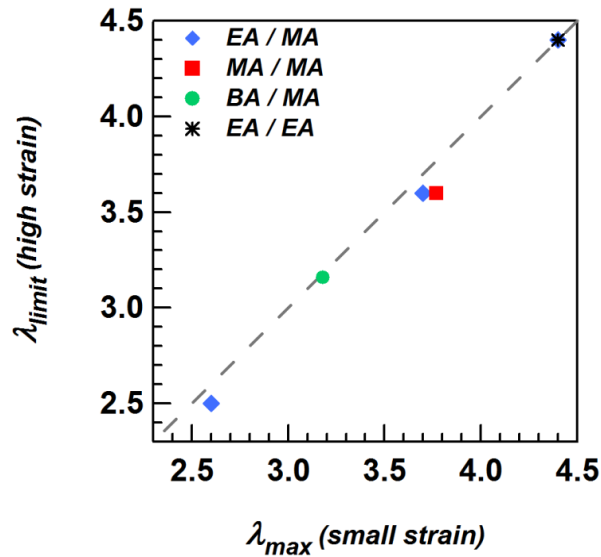

**Figure 11**: Correlation between $\lambda_{limit}$ from fitting **Equation (7)** to the experimental master curve for DN and $\lambda_{max}$ calculated from the maximal extensibility of chains of the first network through the initial modulus at small strain (**Equation (8)**).

**Table 1**: Composition of first networks prepared in solvent, [a][BDA]= $n_{BDA}/n_{monomer}$, with n and $n_{BDA}$ the number of moles of monomer and BDA respectively, [b] $M_x^{th}$ the theoretical weight between crosslinks as $M_x^{th} = (M_0 * n)/(2 * n_{BDA})$, with $M_0$ the molar mass of the monomer.

| Sample | Monomer | [a][BDA] [mol %] | [b]$M_x^{th}$ [g/mol] | Monomer [g] | Toluene [g] | BDA [µL] | HMP [µL] |
|---|---|---|---|---|---|---|---|
| $EA_{05}$ | EA | 1.45 | 3400 | 8.6 | 8.6 | 235.8 | 152.5 |
| $EA_1$ | EA | 2.81 | 1700 | 8.6 | 8.6 | 471.5 | 152.5 |
| $EA_2$ | EA | 5.81 | 860 | 8.6 | 8.6 | 943 | 152.5 |
| $MA_1$ | MA | 2.50 | 1700 | 8.6 | 8.6 | 471.5 | 152.5 |
| $BA_1$ | BA | 3.72 | 1700 | 8.6 | 8.6 | 471.5 | 152.5 |
| $MA_{bulk}$ | MA | 2.5 | 1700 | 8.6 | 0 | 471.5 | 152.5 |

**Table 2**: Formulation of the second network swelling bath, [a][BDA]= $n_{BDA}/n$, with n and $n_{BDA}$ the number of moles of monomer and BDA respectively, [b][HMP]= $n_{HMP}/n$, with $n_{HMP}$ the number of moles of HMP, [c] $M_x^{th}$ the theoretical weight between crosslinks as $M_x^{th} = (M_0 * n)/(2*n_{BDA})$, with $M_0$ the molar mass of the monomer.

| Second monomer | [a][BDA] (mol %) | [b][HMP] (mol %) | [c]$M_c^{th}$ (g/mol) | Monomer (g) | BDA (µL) | HMP (µL) |
|---|---|---|---|---|---|---|
| MA | 0.01 | 0.01 | $4.3 \cdot 10^5$ | 40 | 8.74 | 7.08 |
| EA | 0.01 | 0.01 | $5.0 \cdot 10^5$ | 40 | 7.52 | 6.08 |

**Table 3**: Composition and thermo-mechanical properties of multiple networks with $EA_x$ as first networks, $tan\delta$ and $E$ small strain modulus in MPa measured by DMA; *at 60 °C and ** at 20 °C

| Sample | $\phi_{wt}^{DN}$ | $T_g$ (°C) | $tan\delta$ | $E$ |
|---|---|---|---|---|
| $EA_{05}MA$ | 20 wt % | 13 | 0.05* | 1.5* |
| $EA_1MA$ | 30 wt % | 13 | 0.03* | 2.1* |
| $EA_2MA$ | 35 wt % | 12 | 0.02* | 2.9* |
| $MA_1MA$ | 30 wt % | 22 | 0.08* | 1.7* |
| $BA_1MA$ | 30 wt % | -27 / 15 | 0.05* | 1.9* |
| $EA_{05}EA$ | 20 wt % | -15 | 0.25** | 1.1** |
| $EA_{05}MAMA$ | 6 wt % | 18 | 0.05* | 2.4* |
| $EA_1MAMA$ | 10 wt % | - | - | - |
| $EA_{05}EAEA$ | 6 wt % | -15 | 0.18** | 1.7** |

**Table 4**: Elastic modulus $E$ of simple networks measured from uniaxial elongation and experimental average weight between crosslinks, $M_x^{exp}$, derived from the classical rubber elasticity

| Sample | $E$ (MPa) | $M_x^{exp}$ (g/mol) |
|---|---|---|
| $EA_{05}$ | 0.8 | 10800 |
| $EA_1$ | 1.35 | 7400 |
| $EA_2$ | 2.3 | 3800 |
| $MA_1$ | 1.22 | 7600 |
| $BA_1$ | 1.17 | 7500 |

**Table 5**: Fit parameters for **Equation 7** for the three families of multiple networks, $\lambda_{\text{limit}}$ from fits; $\lambda_{\text{max}}$ from small strain properties of the first network

| 1st Network | DN $\lambda_0$ | TN $\lambda_0$ | DN $\sigma_p$ | TN $\sigma_p$ | $C_1$ | $C_2$ | $N_k$ | $\lambda_{limit}$ | $\lambda_{max}$ |
|---|---|---|---|---|---|---|---|---|---|
| EA$_{0.5}$ | 1.7 | 2.5 | 0.23 | 0.65 | 0.078 | 0.057 | 19.5 | 4.4 | 4.4 |
| EA$_1$ | 1.6 | 2.2 | 0.5 | 0.9 | 0.140 | 0.078 | 13.2 | 3.6 | 3.7 |
| EA$_2$ | 1.3 | - | 0.25 | - | 0.15 | 0.197 | 6 | 2.5 | 2.6 |

**Different brittle and prestretched elastomer networks are embedded at a low volume fraction in a soft extensible matrix.** The increase in toughness of the final material is directly controlled by the non linear elastic properties of the prestretched network and its volume fraction, providing a general design rule for tough soft materials.

### Keyword

elastomers, mechanical properties, double network, fracture, non-linear elasticity

## Characterizing Large Strain Elasticity of Brittle Elastomeric Networks by Embedding them in a Soft Extensible Matrix

### ToC figure

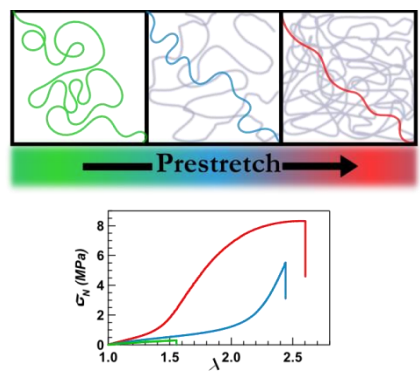

# Supporting Information

**Characterizing Large Strain Elasticity of Brittle Elastomeric Networks by Embedding them in a Soft Extensible Matrix**

*Etienne Ducrot[‡], Costantino Creton\**

Dr. Etienne DUCROT, Prof. Costantino CRETON

Laboratoire Sciences et Ingénierie de la Matière Molle, ESPCI ParisTech, 10 rue Vauquelin, 75231 Paris Cedex 05, FRANCE

E-mail: costantino.creton@espci.fr

[‡]current address CSMR, NYU Physics, 4 Washington pl. 10003 New York, NY

**Transfer reaction, additional crosslink and links between networks**

Two samples of the second network of MA alone were prepared with (MA$_{2N}$) or without (MA$_{2N\_noX}$) crosslinker everything else kept identical.

Those two materials were tested in uniaxial extension, stress strain curves are presented on **Figure S1**. There is not much difference between the two compositions. MA$_{2N\_noX}$ shows nearly the same properties as those of MA$_{2N}$. Using the Rubinstein and Panyukov equation[25] (**Equation S1**), we estimated the contribution to the modulus from entanglements and from crosslinks for the two samples (**Table S1**).

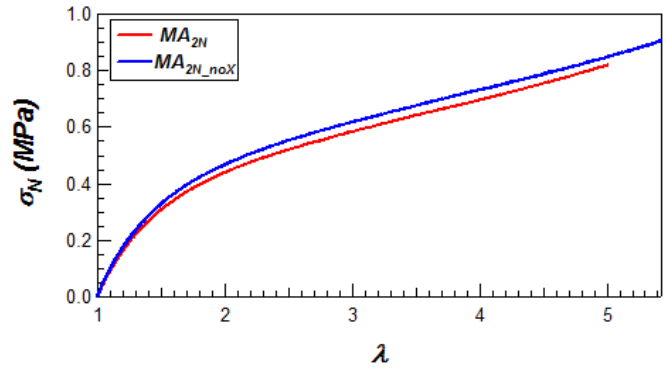

**Figure S1**: Second network alone MA$_{2N}$, with (in red) or without (in blue) BDA as crosslinker

$$\frac{\sigma_N}{\lambda - \frac{1}{\lambda^2}} = G_x + \frac{G_e}{0.74\lambda + 0.61\lambda^{-1/2} - 0.35} \quad \text{(S1)}$$

$$E_i = 3 \cdot G_i$$

The best fits of the uniaxial elongation data to **Equation S1** give the same contribution to the modulus $G_x$ for the two samples. For a sample prepared with no crosslinker, $G_x$ should have been equal to zero and the resulting material would have been a polymer melt. This is evidence that there are transfer reactions during the polymerization which create additional crosslinks.

Assuming rubber-like elasticity we estimated the real crosslinks concentration composed of BDA and transfer reaction of concentration respectively [BDA] and [Tr] (**Equation S2**). The concentration of BDA being set by the starting solution, [Tr] can be estimated for the two samples around 0.1 mol % of monomer (**Table S1**). We have assumed here that $E_x = 3G_x$ from incompressibility.

$$E_x = \frac{3.\rho RT}{M_{mono}.100}([BDA] + [Tr]) \tag{S2}$$

**Table S1**: Properties of second networks of PMA: $MA_{2N}$ and $MA_{2N\_NOX}$ at 60 °C, with $M_{mono}$=86.09 g.mol$^{-1}$, $\rho$=1.12 g.cm$^{-3}$

| Sample | [BDA] | E | $E_e$ | $E_x$ | [Bda]+[Tr] | [Tr] |
|---|---|---|---|---|---|---|
| $MA_{2N}$ | 0.01 mol % | 0.9 MPa | 0.78 MPa | 0.25 MPa | 0.11 mol % | 0.1 mol % |
| $MA_{2N\_noX}$ | 0 mol % | 0.96 MPa | 0.85 MPa | 0.25 MPa | 0.11 mol % | 0.11 mol % |